\documentclass{elsart}
\usepackage{hyperref}

\usepackage{graphicx}
\usepackage{amssymb}

\begin{document}

\begin{frontmatter}


\title{Understanding correlated electron systems by a classification of Mott insulators}
\author{Subir Sachdev}
\address{Department of Physics, Yale University,\\ P.O. Box 208120,
New Haven, CT 06520-8120, USA}
\ead{subir.sachdev@yale.edu}
\ead[url]{http://pantheon.yale.edu/\~\/subir}

\begin{abstract}
This article surveys the physics of systems proximate to Mott
insulators, and presents a classification using conventional and
topological order parameters. This classification offers a
valuable perspective on a variety of conducting correlated
electron systems, from the cuprate superconductors to the heavy
fermion compounds. Connections are drawn, and distinctions made,
between collinear/non-collinear magnetic order, bond order,
neutral spin 1/2 excitations in insulators, electron Fermi
surfaces which violate Luttinger's theorem, fractionalization of
the electron, and the fractionalization of bosonic collective
modes. Two distinct categories of $Z_2$ gauge theories are used to
describe the interplay of these orders. Experimental implications
for the cuprates are briefly noted, but these appear in more
detail in a companion review article (S. Sachdev,
\href{http://arxiv.org/abs/cond-mat/0211005}{cond-mat/0211005}).
\end{abstract}


\end{frontmatter}

\section{Introduction}
\label{sec:intro}

The foundations of solid state physics reside on a few simple
paradigms of electron behavior which have been successfully
applied and extended in a wide variety of physical contexts. The
paradigms include the independent electron theory of Bloch, its
more sophisticated formulation in Landau's Fermi liquid theory,
and the Bardeen-Cooper-Schrieffer (BCS) theory of electron pairing
by an instability of the Fermi surface under attractive
interactions between the electrons.

In the past decade, it has become increasingly clear that these
paradigms are not particularly useful in understanding correlated
electron systems such as the cuprate superconductors and the
``heavy fermion'' compounds. Many electrical and magnetic
properties of these materials are rather far removed from those of
the Fermi liquid, and remain poorly understood despite much
theoretical work. The strong correlations between the electrons
clearly makes the Fermi liquid an inappropriate starting point for
a physical understanding of the many electron ground state.

One approach to the strong correlation problem, advocated here, is
to begin at the point where the breakdown of the Bloch theory is
complete: in the Mott insulator. These are materials in which
Bloch theory predicts metallic behavior due to the presence of
partially filled bands. However, the strong Coulomb repulsion
between the electrons leads to dramatically different insulating
behavior, often with a very large activation energy towards
conduction. A key ingredient in our discussion here will be a
classification of such Mott insulators: we will characterize the
ground states of Mott insulators with a variety of ``order
parameters''. Some of these orders will have a conventional
association with a symmetry of the Hamiltonian which is broken in
the ground state, but others are linked to a more subtle
`topological' order \cite{thoulessbook}.

An implicit assumption in our approach is that these same order
parameters, or their closely related extensions, can also be used
fruitfully in a description of other correlated systems, which may
be either metallic or superconducting. In the latter systems, it
is sometimes the case that the true long-range order characterizes
only a proximate Mott insulator, and the order parameter
`fluctuates' at intermediate scales. In such a situation, the
ground state and its elementary excitations are adiabatically
connected to the Bloch/BCS states, but the order of the proximate
Mott insulator is nevertheless important in understanding
experiments, and especially those that explore correlations at
mesoscopic scales. The theory of quantum phase transitions offers
a powerful method for controlled predictions for such experiments:
identify a quantum critical point associated with the onset of
long-range order, and use it to expand back into the region with
fluctuating order. The reader is referred to Ref.~\cite{rmp} for
the application of such an approach to recent experiments in the
cuprates.

With this motivation, let us turn to the central problem of the
classification of Mott insulators. Newcomers to the Mott insulator
problem should consult Section III of Ref.~\cite{rmp} at this
point for an elementary introduction to the microscopic physics.
The most important degrees of freedom in Mott insulators are the
quantum spins $S_{\alpha} ({\bf r})$, $\alpha = x,y,z$, which
resides on the sites, ${\bf r}$, of some lattice. The states on
each site transform under the spin $S$ representation of SU(2) (we
are usually interested in $S=1/2$), and the sites are coupled
together with the Hamiltonian
\begin{equation}
H = \sum_{{\bf r} < {\bf r}^{\prime}} J({\bf r}, {\bf r}^{\prime}
) S_{\alpha} ({\bf r})S_{\alpha} ({\bf r}^{\prime}) + \ldots,
\label{h}
\end{equation}
where $J({\bf r}, {\bf r}^{\prime} )$ are short-range exchange
interactions (usually all positive, realizing antiferromagnetic
exchange), and the ellipses represent possible multiple spin
couplings, all of which preserve full SU(2) spin rotation
invariance. Here, and henceforth, there is an implied summation
over repeated spin indices.

At the outset, it useful to distinguish Mott insulators by whether
their ground states break the SU(2) spin rotation invariance of
the Hamiltonian or not. The paramagnetic states have $\langle
S_\alpha ({\bf r} ) \rangle = 0$ and preserve SU(2) spin rotation
invariance, while the magnetically ordered states we consider
break spin rotation invariance in the following simple manner
\begin{equation}
\langle S_{\alpha} ({\bf r}) \rangle = {\rm Re} \left[
\Phi_{\alpha} e^{i {\bf K} \cdot {\bf r}} \right]; \label{e1}
\end{equation}
here ${\bf K}$ is the ordering wavevector, while $\Phi_{\alpha}$
is a three-component complex order parameter; this order parameter
transforms as a vector under spin rotations, while $\Phi_{\alpha}
\rightarrow e^{-i {\bf K} \cdot {\bf a}} \Phi_{\alpha}$ under
translations by a Bravais lattice vector ${\bf a}$. For simplicity
we only consider systems with a single ordering wavevector,
although the generalization to multiple wavevectors is not
difficult.

The quantum theory of the magnetically ordered states has been
well established for a long time: one considers slowly varying
quantum fluctuations of the field $\Phi_{\alpha}$ in spacetime,
and so obtains the Goldstone spin-wave excitations. Here, we wish
to push the logic of this approach further: what happens when the
quantum spin-wave fluctuations become strong enough to destroy the
magnetic order in the ground state, and we reach a paramagnetic
state with $\langle \Phi_{\alpha} \rangle = 0$ ? We can think of
this paramagnetic state as one with a `fluctuating'
$\Phi_{\alpha}$ order: does this tell us anything about the
physical properties of the paramagnetic state ? We argue here that
a surprising amount of information can gleaned from this seemingly
naive approach, and it ultimately shows the way to a
classification of both the magnetically ordered and the
paramagnetic states.

\subsection{Landau theory}
\label{sec:landau}

We begin with the simple canonical procedure of considering the
effective potential for $\Phi_{\alpha}$ fluctuations; our
procedure here is general enough that it applies equally to both
quantum and thermal fluctuations in insulators, metals, and
superconductors. The structure of this effective potential is
constrained by spin-rotation invariance and Bravais lattice
translational symmetries, and the following low order terms are
always allowed:
\begin{equation}
V (\Phi_{\alpha}) = s  \Phi_{ \alpha}^{\ast} \Phi_{ \alpha} +
\frac{u}{2} \Phi_{ \alpha}^{\ast} \Phi_{ \beta}^{\ast} \Phi_{
\alpha} \Phi_{\beta}  + \frac{v}{2} \Phi_{ \alpha}^{\ast} \Phi_{
\alpha}^{\ast} \Phi_{ \beta} \Phi_{ \beta}  \label{vphi}
\end{equation}
here $s$, $u$, $v$ are phenomenological Landau parameters ($u>0$,
$v>-u$). If the value of ${\bf K}$ is commensurate with a
reciprocal lattice vector, then additional low order terms can
appear in the effective potential, but we defer considerations of
such terms to later in this subsection.

As is usual, we begin with a minimization of $V$ over the values
of the 3 complex numbers $\Phi_{\alpha}$. For $s>0$, we obtain the
optimum value $\Phi_{\alpha}=0$, which obviously corresponds to
the paramagnet. For $s<0$, we obtain two distinct classes of
minima, which are {\em not} related to each other by any symmetry
of the Hamiltonian. They are
\begin{eqnarray}
&& ({\rm A}) ~\mbox{\bf Non-collinear spins},~~v>0~~:~~\Phi_{
\alpha} = n_{1\alpha} + i n_{2 \alpha} \nonumber \\ &&
~~~~~~~~~~~~~~~~~~~\mbox{with $n_{1,2\alpha}$ real,
 $n_{1 \alpha}
n_{2 \alpha} = 0$ and } n_{1 \alpha}^2 = n_{2 \alpha}^2 =
\frac{-s}{2u}. \nonumber \\
 && ({\rm B}) ~\mbox{\bf Collinear
spins},~~-u<v<0~~:~~\Phi_{ \alpha} = e^{i \theta} n_{\alpha}
\nonumber
\\ &&~~~~~~~~~~~~~~~~~~~~~~~~~~~~~~~ \mbox{with $n_{ \alpha}$ real
and }n_{\alpha}^2 = \frac{-s}{u+v}. \label{ab}
\end{eqnarray}
From (\ref{e1}) we can see easily that in case (B) the average
values of the spins at all ${\bf r}$ are either parallel or
antiparallel to each other, while in (A) the average spins values
map out a circular spiral.

All the solutions in (A) or (B) represent physically distinct
magnetically ordered ground states, but the states within a
category are degenerate and can be transformed to each other by
symmetries of the Hamiltonian. It is useful to more carefully
specify the manifold of degenerate magnetically ordered ground
states. For (A), the ground state manifold is easier to decipher
by a different parameterization of $n_{1,2\alpha}$ which solves
the constraints in (\ref{ab}):
\begin{equation}
n_{1\alpha} + i n_{2 \alpha} = \epsilon_{ac} z_c
\sigma^{\alpha}_{ab} z_b , \label{nz}
\end{equation}
where $a,b$ extend over $\uparrow,\downarrow$, $\sigma^{\alpha}$
are the Pauli matrices, $\epsilon_{ab}$ is the antisymmetric
tensor, and $z_a$ is a two-component complex field with
$|z_\uparrow|^2 + |z_\downarrow|^2 = \sqrt{-s/2u}$; note that
$z_{\alpha}$ transforms like a $S=1/2$ spinor under spin
rotations, while
\begin{equation}
z_{a} \rightarrow e^{-i {\bf K} \cdot {\bf a}/2} z_a, \label{za}
\end{equation}
under translation by a Bravais lattice vector ${\bf a}$. It is
easy to check that (\ref{nz}) is in fact the most general solution
of the constraints for case (A) in (\ref{ab}), but with a two-fold
redundancy: $z_a$ and $-z_a$ correspond to the {\em same}
non-collinear ground state. The two complex numbers $z_a$ are
equivalent to four real numbers, and hence the manifold of ground
states is $S_3 /Z_2$, where $S_3$ is the three-dimensional surface
of a sphere in four dimensions. Turning to (B), $n_{\alpha}$ maps
out the surface of an ordinary sphere, $S_2$, while the phase
factor $e^{i\theta}$ is U(1)$\cong S_1$, a circle. However the
factorization of $\Phi_{\alpha}$ into $n_{\alpha}$ and $e^{i
\theta}$ is redundant because we can map $n_{\alpha} \rightarrow
-n_{\alpha}$ and $\theta \rightarrow \theta + \pi$ without
changing $\Phi_{\alpha}$; hence the manifold of ground states for
(II) is $(S_2 \times S_1)/Z_2$. Summarizing, we have
\begin{eqnarray}
&& ({\rm A}) ~\mbox{\bf Non-collinear spins},~~\mbox{ground state
manifold = } S_3 /Z_2 ; \nonumber \\
&& ({\rm B}) ~\mbox{\bf Collinear spins},~~\mbox{ground state
manifold = } (S_2 \times S_1)/Z_2 . \label{manifold}
\end{eqnarray}

This is a good point to mention additional restrictions that are
placed on the ground state manifold at special commensurate values
of ${\bf K}$. These arise for ${\bf K}$ such that $2p {\bf K}$
equals a reciprocal lattice vector, where $p$ is an integer. Then
from the transformation of $\Phi_{\alpha}$ under Bravais lattice
translations we observe that the effective potential can contain
the additional term
\begin{equation}
V_{\rm comm} = - w \left(\Phi_{\alpha} \Phi_{\alpha} \right)^p +
\mbox{c.c.} \label{comm}
\end{equation}
For case (B) it is easy to see from (\ref{ab}) and (\ref{comm})
that $V_{\rm comm}$ acts only on the angular field $\theta$, and
selects $2p$ values of $\theta$ in the ground state. So the order
parameter manifold is now reduced to $(S_2 \times Z_{2p})/Z_2$:
this change in the manifold has important consequences for $p=1$,
but is not of particular importance at larger values of $p$. For
case (A), it follows from (\ref{ab}) and (\ref{comm}) that $V_{\rm
comm}$ has no influence on the ground state energy, and that the
order parameter manifold remains $S_3/Z_2$.

Having laid the ground work, and now are ready to extend this
simple understanding of magnetically ordered states to
paramagnetic phases, where the order parameter $\Phi_{\alpha}$,
constrained as in (\ref{ab}), is fluctuating. As we will see,
global aspects of the order parameter manifold in (\ref{manifold})
will play a crucial role. We will consider case (A) with
non-collinear spins in Section~\ref{sec:nc}, while case (B) with
collinear spins will be discussed in Section~\ref{sec:collinear}.
In both cases we will attempt to understand a variety of phases,
including insulators, metals, and superconductors. The evidence so
far indicates that the cuprate superconductors are in category
(B), and this is discussed at some length in Ref.~\cite{rmp}. Some
of the more exotic phases appear in category (A), and we
anticipate these will find realizations in heavy-fermion
compounds, and particularly in those in which the magnetic moments
reside on frustrated lattices such as the pyrochlore or the
triangular.

\section{Noncollinear spins}
\label{sec:nc}

An understanding of $\Phi_{\alpha}$ fluctuations in the
paramagnetic phase requires that we proceed beyond the simple
effective potential in (\ref{vphi}), and consider spacetime
dependent fluctuations using a suitable effective action. We
assume that the non-collinearity of the spin correlations is
imposed at some short scale, and so at longer scales we wish to
impose the constraints under (A) in (\ref{ab}) at the outset in
our effective action. This is most conveniently done using the
parameterization in (\ref{nz}). We discretize spacetime on some
regular lattice of sites $j$ on which we have the spinors
$z_{ja}$: note that this lattice may have little to do with the
lattice of sites on which the underlying spins reside, and that we
are working here on a coarse-grained scale. Any effective action
on this coarse-grained lattice must be invariant under global spin
rotations, and also under the global lattice transformation
(\ref{za}). However, most importantly, we note from (\ref{nz})
that it must also be invariant under the $Z_2$ {\em gauge}
transformation
\begin{equation}
z_{ja} \rightarrow \varepsilon_j z_{ja} \label{gauge}
\end{equation}
where $\varepsilon_j = \pm 1$ is a spacetime dependent field which
generates the gauge transformation. This transformation is
permitted because the local physics can only depend upon the order
parameter $\Phi_{\alpha}$, which is invariant under (\ref{gauge}).

We begin in Section~\ref{sec:eff} by considering a very simple
model of $z_{ja}$ fluctuations. This model is surely an
oversimplification for the complex quantum systems of interest
here, but it will at least allow us to obtain an initial
understanding of possible phases and the global structure of the
phase diagram. We will discuss applications to realistic physical
systems in Section~\ref{sec:app}.

\subsection{Simple effective action}
\label{sec:eff}

Our model here omits long-range interactions and mobile
excitations that can carry charge, which implies that it could
apply directly only to insulators. We also neglect all Berry
phases associated with the underlying spins. This will allow us to
at least obtain a first understanding of the possible phases and
the global structure of the possible phase diagrams. More careful
considerations, described partly in Section~\ref{sec:berry} below,
show that the Berry phases are crucial for realizing that `bond
order' is present in the `confining' phase to be discussed
shortly, but that they can be safely neglected in the other
phases. The presentation below is based upon results contained in
Refs.~\cite{rs91,sr91,css94,css94a}.

With the above caveats, we initially introduce the following
simple partition function:
\begin{eqnarray}
\mathcal{Z}_z &=& \int \prod_j d z_{ja} d z^{\ast}_{ja} \delta
\left( \left| z_{ja} \right|^2 - 1 \right) \exp
\left(-\mathcal{S}_z \right) \nonumber \\
 \mathcal{S}_z &=& -\widetilde{J}_1
\sum_{\langle ij \rangle} z^{\ast}_{ia} z_{ib} z^{\ast}_{jb}
z_{ja}-\widetilde{J}_2 \sum_{\langle ij \rangle}
\left(z^{\ast}_{ia} z^{\ast}_{ib} z_{ja} z_{jb} + \mbox{c.c.}
\right) + \ldots, \label{sz}
\end{eqnarray}
where we have rescaled the $z_{ja}$ fields so that they obey the
unit length constraint on every site $j$, and $\langle ij \rangle$
represents nearest neighbors. The terms shown in (\ref{sz})
constitute the most general coupling between two sites consistent
with the symmetries discussed above. Note that at this order the
discrete symmetry (\ref{za}) has effectively restricted us to
terms which are also invariant under the U(1) symmetry $z_{ja}
\rightarrow e^{i \varphi} z_{ja}$ with $\varphi$ arbitrary: so the
model $\mathcal{S}_z$ has a global SU(2)$\times$U(1) symmetry.

While it is possible to proceed with (\ref{sz}), the full spectrum
of possible phases is seen more easily by a representation which
makes the $Z_2$ gauge invariance more explicit. For this, we must
introduce a $Z_2$ gauge field $\sigma_{ij} = \pm 1$, which resides
on the links of the lattice, and which obeys the mapping
\begin{equation}
\sigma_{ij} \rightarrow \varepsilon_i \sigma_{ij} \varepsilon_j
\end{equation}
under the gauge transformation in (\ref{gauge}). We can loosely
view $\sigma_{ij}$ as arising from a Hubbard-Stratonovich
decoupling of the quartic terms in (\ref{sz}). With $\sigma_{ij}$
in hand, we can now propose the alternative action as $Z_2$ gauge
theory
\begin{eqnarray}
\mathcal{Z}_{\rm noncoll} &=& \int \prod_j d z_{ja} d
z^{\ast}_{ja} \delta \left( \left| z_{ja} \right|^2 - 1 \right)
\sum_{\{\sigma_{ij} = \pm 1\}} \exp
\left(-\mathcal{S}_{\rm noncoll} \right) \nonumber \\
 \mathcal{S}_{\rm noncoll} &=& -J_1
\sum_{\langle ij \rangle}  z_{ia}^{\ast} \sigma_{ij} z_{ja} +
\mbox{c.c.} - K \sum_{\Box} \prod_{\langle ij \rangle \in \Box}
\sigma_{ij} -J_2 \sum_{\langle ij \rangle} z^{\ast}_{ia} z_{ib}
z^{\ast}_{jb} z_{ja}. \label{sz2}
\end{eqnarray}
The second term is the sum of the products of $\sigma_{ij}$ over
all elementary plaquettes of the lattice---it is the standard
Maxwell term for a $Z_2$ gauge field. The first term proportional
to $J_1$ is easily seen to have a global O(4) symmetry of
rotations on the order parameter manifold $S_3$. This symmetry is
not present in the underlying spin model, and so we have added the
last term proportional to $J_2$ which reduces the symmetry down to
the required SU(2)$\times$U(1). For $K=0$, we can freely sum over
the $\sigma_{ij}$ independently on each link, and the resulting
action is easily seen to have a structure identical to
$\mathcal{S}_z$ in (\ref{sz}). For large $K$, the action
(\ref{sz2}) will allow us to easily access states which would have
been harder to extract from (\ref{sz}).

\begin{figure}
\centerline{\includegraphics[width=4in]{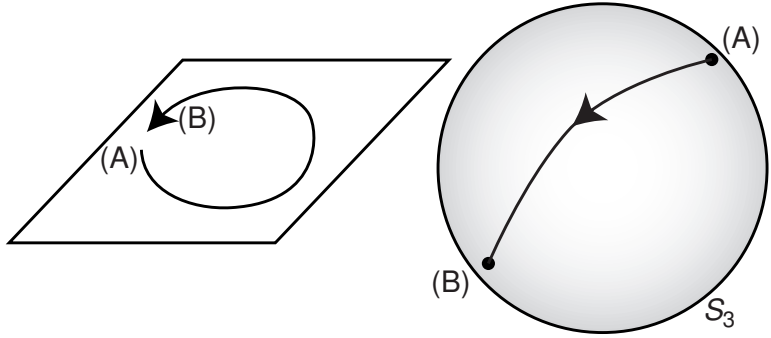}} \caption{A
vison defect associated with $\pi_1 (S_3/Z_2) = Z_2$. The left
hand side represents a two-dimensional section of spacetime, while
the right hand side is the space of points $z_a$ which reside on
$S_3$. Polar opposite points on $S_3$ are represented by A and B.
The $Z_2$ gauge field $\sigma_{ij}=1$ along the loop, but
$\sigma_{ij}=-1$ on the link directly connecting A and B. There is
a $Z_2$ gauge flux at the center of the vison, and this flux
allows identification of the vison defect even when the $z_a$ are
fluctuating, and there is no magnetic order.} \label{fig1}
\end{figure}

What is the physical significance of the flux in the $Z_2$ gauge
field which is controlled by the coupling $K$ in (\ref{sz2}) ? It
is a measure of the location of defects associated with the
homotopy group $\pi_1 (S_3/Z_2) = Z_2$, now often called `visons'
\cite{sf00}. Upon encircling a vison in a closed loop, the values
of $z_a$ change smoothly along the loop, but the initial and final
values of $z_a$ reside on polar opposite points on $S_3$; all
physical energies are independent of this sign change, and so
there is no practical significance to this cut (see
Fig~\ref{fig1}).
From (\ref{sz2}) we see that the action will be minimized by
$\sigma_{ij} = 1$ along the loop, but we prefer $\sigma_{ij} = -1$
across the cut between the initial and final points . Measuring
the $Z_2$ flux associated with this configuration of
$\sigma_{ij}$, we conclude that the flux resides in a small region
at the center of the vison.

\begin{figure}
\centerline{\includegraphics[width=4in]{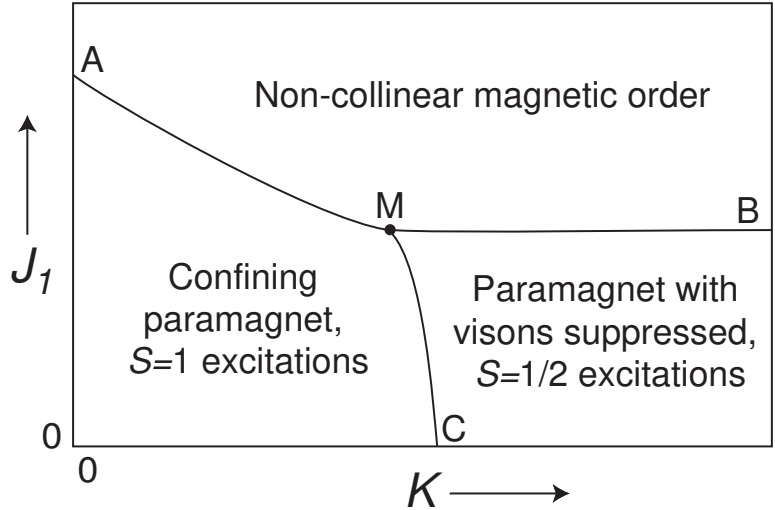}}
\caption{Schematic phase diagram of the model $\mathcal{S}_{\rm
noncoll}$ in (\protect\ref{sz2}) as a function of $J_1$ and $K$ in
2+1 spacetime dimensions. We assume that $J_2$ takes a small
positive value. The model $\mathcal{S}_{\rm noncoll}$ describes
some of the physics of Mott insulators with non-collinear spin
correations. However it omits Berry phase terms which impose the
Heisenberg commutation relations of the spins in
(\protect\ref{h}). These Berry phases play an important role in
the small $K$ region of the phase diagram: they induce bond order
in the confining paramagnet, along with additional intermediate
phases. The influence of the Berry phases is studied in
Section~\protect\ref{sec:berry} and in Ref.~\protect\cite{sp02}.
The suppression of visons in the deconfined state at large $K$ is
interpreted as a `topological' order.} \label{fig2}
\end{figure}

We can now sketch a phase diagram of $\mathcal{S}_{\rm noncoll}$
in the $J_1,K$ plane in 2+1 spacetime dimensions (related
considerations apply to higher spacetime dimensions). Many
important features follow immediately by an analogy with a simpler
problem considered by Lammert {\em et al.}
\cite{lammert93,lammert95} in the entirely different context of
liquid crystals: they examined a problem with an order parameter
belonging to the space $S_2/Z_2$ (in contrast to the $S_3/Z_2$
order parameter of interest here), which also permits an effective
action in a $Z_2$ gauge theory very similar to $\mathcal{S}_{\rm
noncoll}$. Using their results, and those of the earlier work of
Refs.~\cite{wegner71,fradkin79}, we obtain the phase diagram for
2+1 spacetime dimensions shown in Fig~\ref{fig2}.

Consider first the physics near the line $K=0$. Here we can simply
sum over the $\sigma_{ij}$, and work instead with the original
action $\mathcal{S}_z$ in (\ref{sz}). At large $J_1$ we have a
conventional non-collinear magnetically ordered phase with
$\langle z_a \rangle \neq 0$, and hence $\langle \Phi_{\alpha}
\rangle \neq 0$. With decreasing $J_1$, there is a phase
transition to a paramagetic phase. The key to understanding the
nature of this paramagnetic phase, and of the quantum phase
transition, is the observation that $\mathcal{S}_z$ contains only
gauge-invariant bilinears in $z_{ja}$ on each site, and so we
should be able to describe the physics in terms of the bilinear
fields $n_{1,2\alpha}$ in (\ref{nz}). Alternatively stated, the
strong $Z_2$ gauge fluctuations (and the accompanying
proliferation of visons) has confined the $z_a$ quanta, and the
elementary excitations of the paramagnetic state are the
$n_{1,2\alpha}$ quanta. Notice that these fields carry spin $S=1$,
and so the paramagnetic phase is a confining state with only
integer spin excitations. The quantum transition between the
magnetically ordered state and this confining state, along the
line AM in Fig~\ref{fig2} can be addressed by continuum models
expressed directly in terms of the $n_{1,2\alpha}$. This has been
done by Kawamura \cite{kawamura88}, Pelissetto {\em et al.}
\cite{pelissetto01,pelissetto01a}, and Calabrese {\em et al.}
\cite{calabrese02} in the context of classical phase transitions
in stacked triangular lattices, and they obtained a continuous
phase transition in the ``chiral'' universality class in 2+1
spacetime dimensions. The global symmetry at this ``chiral'' fixed
point remains SU(2)$\times$U(1).

Let us turn now to the physics near the line $J_1=0$ in
Fig~\ref{fig2}. Here, the $z_{a}$ fluctuate strongly and can be
integrated out, leading to a pure $Z_2$ gauge theory in the
$\sigma_{ij}$. In 2+1 spacetime dimensions, this gauge theory is
well known \cite{wegner71,fradkin79} to have a
confinement-deconfinement transition at some critical $K$ near the
point C: the large $K$ region is the deconfined state where vison
fluxes are suppressed. Consequently, in this deconfined state, we
can use a simple physical picture in which we choose a gauge with
all $\sigma_{ij}=1$: the $z_a$ quanta in (\ref{sz2}) are then free
to propagate through the system. The large $K$, small $J_1$ region
is therefore a paramagnetic state with neutral $S=1/2$ `spinon'
excitations. The suppression of visons in this state is
interpreted as the presence of `topological order'; this is
similar to the topological order below the Kosterlitz-Thouless
transition in the two-dimensional classical XY model, where point
vortices are suppressed \cite{thoulessbook}.

Finally, to complete our picture of the phase diagram, we look at
the region of large $K$. Here, visons are suppressed, and so we
can freeze $\sigma_{ij} = 1$, and we are left with a simple model
of interacting $z_{\alpha}$ quanta, with a global
SU(2)$\times$U(1) symmetry. Such a theory has been studied in its
continuum limit, and yields a description of the magnetic ordering
transition along the line BM: the critical point has a large O(4)
symmetry in 2+1 spacetime dimensions, and the critical exponents
are those of the 4-component $\varphi^4$ field theory
\cite{azaria90,css94}.

\subsection{Physical Applications}
\label{sec:app}

The results of Section~\ref{sec:eff} for large $K$ apply directly
to Mott insulators with non-collinear spin correlations. The
mapping of the magnetically ordered phase is evident, while we
identify the paramagnet with visons suppressed with the resonating
valence bond (RVB) state of
Refs~\cite{pauling49,fazekas74,Anderson87,krs87,wen91,moessner01}.
The latter identification follows from the full preservation of
symmetries of the Hamiltonian, the presence of neutral $S=1/2$
spinon excitations, and from the topological order implied by the
suppression of visons \cite{rs91,wen91,sf00}. Experimentally, we
note that the evidence for a RVB state in Cs$_2$CuCl$_4$
\cite{coldea01} is in a system with non-collinear spin
correlations, consistent with our theoretical picture. At small
$K$, and smaller $J_1$, the proliferation of visons requires more
care, as the quantum mechanics of the underlying spins contributes
a Berry phase to each vison \cite{jalabert91,sf00}. These phases
induce bond order in the confining paramagnet, and can also lead
to additional intermediate states \cite{sf00,sp02}: we will not
discuss this further here, but will illustrate the influence of
Berry phases in the simpler context of collinear spins in
Section~\ref{sec:berry}.

Let us finally move away from the Mott insulator, and introduce
mobile charge carriers into the RVB state just discussed, by
doping in holes. A scenario which has been intensively discussed
in the literature \cite{Anderson87,krs87,sf00} is that hole
fractionalizes into independent quasiparticle excitations: a
neutral $S=1/2$ `spinon' and a spinless, charge $e$ holon. The
spinon here is, of course, that just discussed above here in the
proximate Mott insulator. We can express this fractionlization of
the injected hole by the following schematic operator relation for
the electron annihilation operator $c_a$:
\begin{equation}
c_{a} = f^{\dagger} \left( \lambda_1 z_a + \lambda_2 \epsilon_{ab}
z_b^{\ast} \right), \label{sc}
\end{equation}
where $z_a$ is the bosonic neutral spinon operator from
Section~\ref{sec:eff}, $f^{\dagger}$ is a fermionic operator that
creates a spinless hole, and $\lambda_{1,2}$ are constants that
depend upon the details of the microscopic physics. It is possible
for the fermionic statistics to pass from the hole to the spinon
by binding between quasiparticles and visons, as has been
discussed in some detail in Refs.~\cite{kivelson89,rc89,demler02}.
Then the relationship (\ref{sc}) would be replaced by
\begin{equation}
c_{a} = b^{\dagger}\left( \lambda_3 f_a + \lambda_4 \epsilon_{ab}
f^{\dagger}_b \right), \label{sc2}
\end{equation}
where $f_a$ is a neutral, $S=1/2$, fermionic spinon, $b$ is charge
$e$ bosonic holon, and $\lambda_{3,4}$ are constants. In this
scenario, the Bose condensation of the holon leads to a BCS
superconductor with vestiges of the topological order of the RVB
Mott insulator: experimental probes of this `fluctuating'
topological order have been proposed \cite{ss92,nl92,sf01a,rmp},
but no positive signal has been observed so far in the cuprates
\cite{wynn01,bonn01}.

\subsubsection{Fractionalized Fermi liquids}
\label{sec:small}

It has been argued recently \cite{ssv02,burdin02} that a {\em
different} possibility is more likely for the doped RVB Mott
insulator in $d\geq 2$ spatial dimensions. The doped electrons (or
holes), instead of fractionalizing into spinless charged particles
and neutral $S=1/2$ spinons, retain their integrity in the ground
state, and their spin and charge remain bound to each other. At
the same time, the neutral $S=1/2$ spinons of the RVB Mott
insulator survive in the doped system, and are only renormalized
slightly by the mobile carriers. So, alternatively stated, the
added electrons form a Fermi-liquid-like state which is
approximately decoupled from the spinons. The resulting metallic
state has a Fermi surface with $S=1/2$, charge $-e$ quasiparticle
excitations, along with a {\em separate\/} set of neutral spinon
excitations which continue from the RVB Mott insulator. Moreover,
the volume enclosed by the Fermi surface is `small' {\em i.e.} it
is determined solely by the density of doped electrons, and does
not include the spins of the Mott insulator. In this situation,
the volume of the Fermi surface becomes a direct experimental
signal of the topological order in the RVB state. Heavy fermion
compounds on frustrated lattices with weak or absent magnetic
order are likely candidates for realizing this state.

The fractionalized Fermi liquid defined above violates the
standard Luttinger theorem, and it is useful to express this
violation in its most general terms. Consider a correlated
electron system on a periodic lattice in $d$ spatial dimensions,
whose ground state preserves time-reversal and spin rotation
invariance, and has unit cell volume $v_0$. Let $n_T$ be the {\em
total} density of electrons per volume $v_0$; $n_T$ includes all
the electrons in the system, including {\em e.g.} the core 1s
electrons. Luttinger's theorem states that in a conventional Fermi
liquid state
\begin{equation}
2 \times \frac{v_0}{(2 \pi)^d} \times \left( \mbox{Volume enclosed
by Fermi surface} \right) = n_T (\mbox{mod 2}). \label{luttinger1}
\end{equation}
The leading factor of 2 on the left hand side comes from spin
degeneracy, while modulus 2 on the right hand side allows fully
filled bands to not contribute to the Fermi surface volume. The
fractionalized Fermi liquid being discussed here violates
(\ref{luttinger1}), but instead obeys a modified relation
\begin{equation}
2 \times \frac{v_0}{(2 \pi)^d} \times \left( \mbox{Volume enclosed
by Fermi surface} \right) = (n_T-1) (\mbox{mod 2}).
\label{luttinger2}
\end{equation}
So exactly one electron per unit cell has dropped out from the
Fermi volume. This can only happen in a topologically ordered
state which possesses neutral $S=1/2$ spinon excitations, in
addition to the electron-like quasiparticle excitations on the
small Fermi surface.

\section{Collinear spins}
\label{sec:collinear}

We turn here to the second broad category of correlated electron
systems introduced in Section~\ref{sec:intro}: the collinear spin
case (B) in (\ref{ab}). We will find that the paramagnetic states
with fluctuating collinear spin order are entirely different from
those present for non-collinear spins in Section~\ref{sec:app}.
The order parameter manifold, from (\ref{manifold}) is now $(S_2
\times S_1)/Z_2$, and the appearance of a $Z_2$ quotient means
that the physics can again be described in a generalized phase
diagram of a $Z_2$ gauge theory \cite{zds02}. However, the $Z_2$
flux now identifies a new type of defect which we will discuss in
Section~\ref{sec:efrac}. The $Z_2$ gauge theory also has a
deconfined phase, but this $Z_2$ fractionalization does not lead
to neutral $S=1/2$ excitations; instead, as we will see in
Section~\ref{sec:efrac} it is the spin and charge collective modes
which `fractionalize' apart from each other.

A separate crucial property of Mott insulators with collinear spin
correlations in spatial dimension $d=2$ is the ubiquity of
confining states with only integer spin excitations \cite{rs89b}.
Moreover, except for certain special values of the spin $S$ per
unit cell, these confining states break lattice space group
symmetries by the appearance of spontaneous `bond order' in the
ground state. This bond order also often survives in proximate
conducting states obtained by doping the Mott insulator
\cite{sr91,vs99}. Let us define bond order more precisely here:
most generally, bond order implies a modulation in observables
invariant under spin rotation and time reversal which break a
space group symmetry of the Hamiltonian. The simplest such
observable in states proximate to Mott insulators is simply the
spin exchange energy: so we can introduce the bond order parameter
\begin{equation}
Q_{\bf a} ({\bf r}) = \langle S_{\alpha} ({\bf r}) S_{\alpha}
({\bf r} + {\bf a} ) \rangle, \label{bond}
\end{equation}
where ${\bf a}$ is usually a vector connecting near neighbors.
Note that for ${\bf a} = 0$, $Q_0 ({\bf r})$ is a measure of the
charge density on site ${\bf r}$, and so (if permitted by the
symmetry of the state) there is usually also a modulation of the
site charge density in a bond-ordered state. However, we expect
the long-range Coulomb interactions to suppress modulations at
${\bf a}=0$, while those with ${\bf a} \neq 0$ should be
significantly larger.

We will begin our discussion in Section~\ref{sec:mottberry} by a
detailed discussion of the simplest, and most common, Mott
insulator with collinear spin correlations: that on a
$d$-dimensional cubic lattice with ordering wavevector ${\bf K} =
(\pi,\pi, \ldots)$. This order is commensurate with $p=1$, in the
notation of (\ref{comm}). We will explicitly evaluate the quantum
spin Berry phases for this case, show their intimate connection to
bond order. Section~\ref{sec:dopeberry} will also include
extensions of the study of bond order to doped systems with mobile
carriers.

States at other wavevectors and in metals and superconductors, and
the $Z_2$ gauge theory of the fractionalization of their
excitations will be discussed in Section~\ref{sec:efrac}.

\subsection{Berry phases and bond order}
\label{sec:berry}

\subsubsection{Mott insulators}
\label{sec:mottberry}

Consider the insulating antiferromagnetic (\ref{h}) on a
$d$-dimensional cubic lattice with predominant nearest-neighbor
exchange interactions. This should have collinear spin
correlations with ${\bf K} = (\pi,\pi,\ldots)$, which allows the
term with $p=1$ in (\ref{comm}). Inserting $\Phi_{\alpha} =
e^{i\theta} n_{\alpha}$ (obtained from (\ref{ab})) into
(\ref{comm}) we observe that we can always choose the origin of
co-ordinates so that the values $\theta = 0, \pi$ are preferred.
These do not lead to new values of $\Phi_{\alpha}$, and so in this
case we have simply $\Phi_{\alpha} = n_{\alpha}$, a real
three-component vector.

Now express the coherent state path integral of (\ref{h}) using
the values of the field $n_{\alpha}$ on a $d+1$ dimensional
hypercubic lattice discretization of spacetime. The derivation of
this path integral is reviewed in Chapter 13 of Ref.~\cite{book},
and leads to the following partition function
\begin{equation}
\mathcal{Z}_n = \int \prod_j d n_{j\alpha} \delta(n_{j\alpha}^2 -
1) \exp \left( - \frac{1}{2g} \sum_{\langle ij \rangle}
n_{i\alpha} { n}_{j\alpha} - i S \sum_j \eta_j \mathcal{A}_{j\tau}
\right), \label{f1}
\end{equation}
where $i,j$ are sites of the $d+1$ dimensional hypercubic lattice
(the symbol $i=\sqrt{-1}$ in the prefactor of the second term, and
the context should prevent confusion on its meaning), and we have
rescaled all the $n_{j \alpha}$ to make them unit length. The
first term in the action above is the analog of the terms in
(\ref{sz}), and imposes a cost in the action for deviations from
the perfectly ordered state; we expect a magnetically ordered
state $\langle n_{\alpha} \rangle \neq 0$ for small values of the
coupling $g$, and a paramagnetic state with $\langle n_{\alpha}
\rangle = 0$ for large $g$. The second term in (\ref{f1}) is the
all important Berry phase: here $\eta_j = e^{i {\bf K} \cdot {\bf
r}_j} = \pm 1$ is a fixed field identifying the {\em spatial}
sublattice of the site $j$ (note that $\eta_j$ is independent of
the imaginary time co-ordinate, $\tau$). Finally $\mathcal{A}_{j
\mu}$, with $\mu = \tau,x,y, \ldots$ taking $d+1$ dimensional
possible values, is defined by
\begin{eqnarray}
\mathcal{A}_{j\mu} &\equiv& \mbox{oriented area enclosed by the
spherical triangle with vertices}\nonumber \\
&~&~\mbox{$n_{j\alpha}$, $n_{j+\hat{\mu},\alpha}$, and any {\em
fixed} reference $N_{0\alpha}$},
\end{eqnarray}
where $j+\hat{\mu}$ is the nearest site to $j$ in the $\mu$
direction. It is customary to choose $N_{0\alpha} = (0,0,1)$, the
north pole, but it is not difficult to see that the partition
function $\mathcal{Z}_n$ is independent of the value of
$N_{0\alpha}$. Indeed, elementary geometric considerations of
spherical triangles show \cite{sp02} that a choice of a different
$N_{0\alpha}^{\prime}$ leads to $\mathcal{A}_{j\mu}^{\prime}$
which is related to $\mathcal{A}_{j\mu}$ by a `gauge
transformation'
\begin{equation}
\mathcal{A}_{j\mu}^{\prime} = \mathcal{A}_{j\mu} - \Delta_{\mu}
\phi_j \label{triangles}
\end{equation}
where $\Delta_{\mu}$ is the discrete lattice derivative in the
$\mu$ direction, and $\phi_j$ is the area of the spherical
triangle formed by $n_{j\alpha}$, $N_{0\alpha}$ and
$N_{0\alpha}^{\prime}$. It should also be noted here that the area
of a spherical triangle is uncertain modulo $4 \pi$, and so is the
value of $\mathcal{A}_{j\mu}$, but the partition function
$\mathcal{Z}_n$ is insensitive to this uncertainty because $e^{i 4
\pi S} = 1$.

For small $g$ in $\mathcal{Z}_n$, fluctuations of $n_{\alpha}$ are
suppressed, and in dimensions $d>1$, we are in the conventional
N\'{e}el ordered ground state with $\langle n_{\alpha} \rangle
\neq 0$. As an aside, we note the small $g$ behavior for $d=1$. In
$d=1$ we can continue to assume that $n_{\alpha}$ varies smoothly
from site-to-site for small $g$, and hence take the naive
continuum limit of (\ref{f1}). This yields the O(3) non-linear
sigma model in 1+1 dimensions, along with a topological
$\theta$-term with co-efficient $\theta = 2 \pi S$, as reviewed in
Ref.~\cite{book}. For integer $S$ this exhibits the Haldane gap
state, while for half-odd-integer $S$ a gapless critical state as
in the Bethe's solution of the nearest neighbor antiferromagnetic
chain can appear.

Our interest here is primarily in the larger $g$ regime of
$\mathcal{Z}_n$, where there are significant fluctuations of
$n_{\alpha}$, and we are in a paramagnet with $\langle n_{\alpha}
\rangle = 0$. There are large fluctuations here in the value of
the $\mathcal{A}_{j \mu}$ also, and so the Berry phase term
requires careful evaluation. An explicit evaluation of this term
is essentially impossible, but considerable progress has been made
in the `easy-plane' case, where spin-anisotropies reduce
fluctuations to the equator in spin space with $n_{j \alpha} =
(\cos\theta_j, \sin \theta_j, 0)$ \cite{sp02}. Here we follow a
simple `hand-waving' argument \cite{altenberg} whose results are
known to be consistent with all cases in which more sophisticated
methods are possible; the paragraph below is partly reproduced
from the review in Ref.~\cite{altenberg}, to which we will also
refer the reader for additional details.

The main idea for the larger $g$ regime is to change variables
from the order parameter $n_{j \alpha}$ to the field
$\mathcal{A}_{j\mu}$ which naturally represents the Berry phase.
Formally, this can be done by introducing new `dummy' variables
$A_{j \mu}$ and rewriting (\ref{f1}) by introducing delta-function
factors which integrate to unity on each link; this leads to
\begin{eqnarray}
\mathcal{Z}_n &=& \int \prod_{j \mu} d A_{j \mu} \exp \left(- i 2S
\sum_j \eta_j A_{j\tau} \right) \int \prod_j d n_{j\alpha}
\delta(n_{j\alpha}^2 - 1) \delta(\mathcal{A}_{j\mu}/2 - A_{j\mu}) \nonumber \\
&~&~~~~~~~~~~~~~~~~~~~~~~~~~~~~~~~~~~~~~~~~\times\exp \left( -
\frac{1}{2g} \sum_{\langle ij \rangle} n_{i\alpha}  {n}_{j\alpha}
\right)
\nonumber \\
&=& \int \prod_{j \mu} d A_{j \mu} \exp \left(-\mathcal{S}_A
(A_{j\mu}) - i 2 S \sum_j \eta_j A_{j\tau} \right).
 \label{f2a}
\end{eqnarray}
In the first expression, if the integral over the $A_{j \mu}$ is
performed first, we trivially return to (\ref{f1}); however in the
second expression we perform the integral over the $n_{j\alpha}$
variables first at the cost of introducing an unknown effective
action $\mathcal{S}_A$ for the $A_{j \mu}$. In principle,
evaluation of $\mathcal{S}_A$ may be performed order-by-order in a
``high temperature'' expansion in $1/g$: we match correlators of
the $A_{j \mu}$ flux with those of the $\mathcal{A}_{j \mu}$ flux
evaluated in the integral over the $n_{j\alpha}$ with positive
weights determined only by the $1/g$ term in (\ref{f1}). Rather
than undertaking this laborious calculation, we can guess
essential features of the effective action $\mathcal{S}_A$ from
some general constraints. First, correlations in the $n_{j\alpha}$
decay exponentially rapidly for large $g$ (with a correlation
length $\sim 1/\ln(g)$), and so $\mathcal{S}_A$ should be local.
Second, (\ref{triangles}) implies that $\mathcal{S}_A$ should be
invariant under the gauge transformation
\begin{equation}
A_{j \mu}^{\prime} = A_{j \mu} - \Delta_{\mu} \phi_j /2.\label{f3}
\end{equation}
Finally, uncertainity of $\mathcal{A}_{j \mu}$ modulo $4 \pi$
implies that $\mathcal{S}_A$ should also be invariant under
\begin{equation}
A_{j \mu} \rightarrow A_{j \mu} + 2 \pi. \label{f4}
\end{equation}
The simplest local action which is invariant under (\ref{f3}) and
(\ref{f4}) is that of {\em compact U(1) quantum `electrodynamics'}
and so we propose that for larger values of $g$, $\mathcal{Z}_n
\approx \mathcal{Z}_A$ with
\begin{equation}
\mathcal{Z}_A = \int \prod_{j\mu} d A_{j \mu} \exp \left(
\frac{1}{e^2} \sum_{\Box} \cos \left(\Delta_{\mu} A_{j \nu } -
\Delta_{\nu} A_{j \mu} \right) - i 2 S \sum_j \eta_j A_{j\tau}
\right); \label{f5}
\end{equation}
comparison with the large $g$ expansion shows that the coupling
$e^2 \sim g^2$. This is one of the central results of this
subsection \cite{rs89b,fradkin90,jalabert90,sp02}: the
paramagnetic state of a Mott insulator with two-sublattice
collinear spin correlations in $d$ spatial dimensions is described
by a compact U(1) gauge theory in $d+1$ spacetime
dimensions\footnote{Some readers may find the following
alternative interpretation of the compact U(1) gauge theory
useful. Express $n_{j\alpha} = w^{\ast}_{ja} \sigma^{\alpha}_{ab}
w_{jb}$ where $w_{ja}$ is a two-component complex spinor. This
parameterization has a U(1) gauge redundancy corresponding to
$w_{ja} \rightarrow e^{i \phi} w_{ja}$ (contrast this with the
$Z_2$ gauge redundancy associated with (\protect\ref{nz})).
Expressed in terms of the $w_{ja}$, the theory is a lattice
discretization of the CP$^1$ model
\protect\cite{dadda78,witten79,sp02}, with $A_{j\mu}$ its compact
U(1) gauge field. Integrating out the $w_{ja}$ in the paramagnetic
phase leads to (\protect\ref{f5}).}, accompanied by a Berry phase
term as specified in (\ref{f5}). In the gauge theory/particle
physics language, the Berry phase corresponds to static `matter'
fields of charge $\pm 2S$ residing on the two spatial sublattices
of the $d$-dimensional lattice.

We are now faced with the technical problem of evaluating the
partition function of compact QED in $d+1$ spacetime dimensions in
the presence of static background matter fields, as written in
(\ref{f5}). There is a good understanding of the phases of this
theory in $d=1,2$, and the reader is referred to another review
\cite{altenberg} for further details---here we will summarize the
main conclusions. The structure in $d=3$ is not yet fully
understood, and this remains an important avenue for future
research. All of these previous analyses rely on a duality mapping
of (\ref{f5}); the mapping proceeds in a canonical manner,
beginning with a rewriting of the cosine term in (\ref{f5}) in the
Villain periodic Gaussian form, followed by the evaluation of the
integral over the $A_{j \mu}$, which results in a partition
function of `dual' fields. The advantage of this duality mapping
is that the Berry phases are exactly accounted for at the outset,
and the partition function in the dual representation has only
positive weights---this makes it amenable to standard analyses of
statistical field theory. A discussion of the results in various
spatial dimensions, $d$, follows.

\begin{figure}
\centerline{\includegraphics[width=4in]{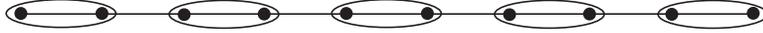}}
\caption{Schematic representation of the ground state of
$\mathcal{Z}_A$ for half-odd-integer $S$ in $d=1$. The state is
identified with the different values of $Q_{\bf a} ({\bf r})$ in
(\protect\ref{bond}) with ${\bf a}$ a nearest neighbor vector: the
links with ellipses have values of $Q_{{\bf a}} ({\bf r})$
distinct from those without. The reader can also view this picture
as caricature of the wavefunction---a simple trial state for
$S=1/2$ is a product of singlet valence bonds between sites linked
by an ellipse} \label{fig3}
\end{figure}

In $d=1$, the duality mapping of the Villain form of
$\mathcal{Z}_A$ leads to a simple partition function that can be
evaluated exactly \cite{altenberg,sp02}. There is a gap to all
excitations, and for half-odd-integer $S$ a two-fold degenerate
ground state and a broken translational symmetry associated with
the appearance of bond order, as illustrated in Fig~\ref{fig3}.

In $d=2$, the duality mapping of (\ref{f5}) leads to {\em
interface} models (also called height or solid-on-solid models) in
2+1 dimensions \cite{rs89b,jalabert90,sp02} (remarkably, the same
interface models are also obtained by a duality mapping
\cite{fradkin90,rs90} on quantum dimer models \cite{rk88} of the
paramagnetic state). In statistical mechanics, these interface
models would describe the growth of a three-dimensional interface
of a four-dimensional crystal. The Berry phases in (\ref{f5}) lead
to offsets in the allowed configurations of the interface model. A
fundamental property of interface models in 2+1 dimensions is that
their configurations are {\em smooth} for all couplings. In the
smooth state, the symmetry of uniform translations of the
interface by a constant is broken, and the interface has some
fixed average height. When combined with the offsets produced by
the Berry phases, the smooth interface is seen to imply a broken
lattice space group symmetry because of the appearance of bond
order (except for certain special values of $S$ which depend upon
the lattice structure---for the square lattice, there is no bond
order for even integer $S$). The fluctuations of the smooth
interface are gapped, and these correspond to a gapped singlet
excitation mode of the quantum antiferromagnet. In the QED
langauge, the gauge-field excitations are gapped, and the compact
U(1) gauge theory is in a confining state. The spinful excitations
of the antiferromagnet consist of a gapped $S=1$ mode associated
with the fluctuations of $n_{\alpha}$---there are no neutral
excitations which carry $S=1/2$.
\begin{figure}
\centerline{\includegraphics[width=4in]{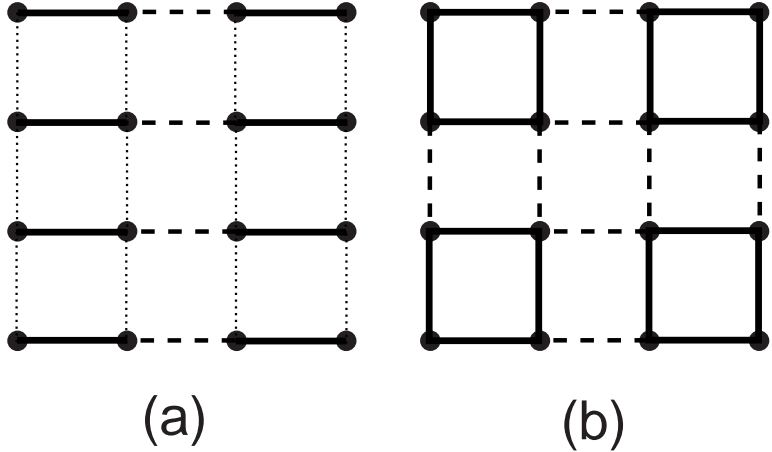}} \caption{Sketch
of the two simplest possible states with bond order for $S=1/2$ on
the square lattice: (a) the columnar spin-Peierls states, and (b)
plaquette state. The different values of $Q_{\bf a} ({\bf r})$,
defined in (\protect\ref{bond}), on the links are encoded by the
different line styles. Both states are 4-fold degenerate; an
8-fold degenerate state, with superposition of the above orders,
also appears as a possible ground state of the generalized
interface model.} \label{fig4}
\end{figure}

A full analysis of the situation in $d=3$ has not been carried
out. However, it is clear that a new state can appear in the phase
diagram--the compact QED theory has a gapless Coulomb phase with a
`photon' excitation \cite{banks77,fradkin78,wen02a,motrunich02}.
In the quantum antiferromagnet, this photon corresponds to a
collective spinless excitation with a gapless linear dispersion.
It is likely that this Coulomb phase is insensitive to the Berry
phases, has no bond order, and allows neutral $S=1/2$ spinon
excitations. The analog of the bond-ordered states of $d=2$ should
also be present.

Large scale numerical studies of $S=1/2$ quantum antiferromagnets
on the square lattice which allow for the destruction of magnetic
order in the ground state have only just become possible. Two
studies have appeared recently \cite{sandvik02,harada02}, and both
present unambiguous evidence of bond order as shown in
Fig~\ref{fig4}.

\subsubsection{Mobile charge carriers}
\label{sec:dopeberry}

The fate of the $d=2$ bond-ordered paramagnet upon doping with
mobile charge carriers has been investigated in a number of
studies \cite{sr91,vs99,vzs00,park01,vojta02}. As stated earlier,
the bond order survives for a finite range of doping
concentrations. Moreover, the ground state also acquires
$d$-wave-like superconducting order which co-exists with bond
order, a state first predicted in Ref.~\cite{sr91}. Loosely
speaking, this superconductivity can be understood as a
consequence of the mobility of the singlet valence bonds in
Fig~\ref{fig3}: the mobile pairs of electrons behave as Cooper
pairs, and their Bose-Einstein condensation leads to
superconductivity. The configuration and period of the bond order
can also evolve with increasing density of carriers
\cite{vs99,vzs00,vojta02}, especially if the parameters are such
that the physics of frustrated phase separation is important
\cite{ekl90}. The reader is referred to Ref.~\cite{rmp} for a
recent review of this work, along with implication for a number of
recent experiments on the cuprate superconductors. These
experiments include a possible observation in STM of a state with
co-existing bond order and $d$-wave superconductivity
\cite{howald02a,howald02b,podolsky02,vojta02,zhang02}.

\subsection{Collective mode fractionalization}
\label{sec:efrac}

We now consider magnetically ordered and paramagnetic states with
collinear spin correlations at wavevectors other than ${\bf K} =
(\pi,\pi, \ldots)$. For these cases the field $e^{i \theta}$, in
the decomposition $\Phi_{\alpha} = e^{i \theta} n_{\alpha}$ in
(\ref{ab}), cannot be disposed at the outset, and needs to be
retained as an additional degree of freedom. The physical
significance of this additional field is readily apparent from a
comparison of the definitions (\ref{e1}) and (\ref{bond}) of the
magnetic order parameters; a factorization of the expectation
values in (\ref{bond}) implies that there is contribution
\cite{zachar98,psv02}:
\begin{equation}
Q_{\bf a} ({\bf r}) = \mbox{Re} \left[ e^{2i \theta} e^{i {\bf K}
\cdot (2 {\bf r} + {\bf a})}\right] + \ldots \label{theta}
\end{equation}
where we have assumed, as in Section~\ref{sec:mottberry}, that
$n_{\alpha}$ has been rescaled so that $n_{\alpha}^2 = 1$. So a
collinear magnetically ordered state at wavevector ${\bf K}$
automatically has co-existing bond order modulations at wavevector
$2 {\bf K}$, and $e^{2 i \theta}$ is the bond order parameter.

We can now combine this observation with the results of
Section~\ref{sec:mottberry} to obtain a simple, and quite general,
theory of the loss of magnetic order for the values of ${\bf K}$
under consideration here. As bond order co-exists with magnetic
order, it is natural consider the possibility that the loss of
magnetic order occurs in a `background' of bond order, and the
latter is present on both sides of the transition. In other words,
we consider a transition from a magnetically ordered state with
$\langle \Phi_{\alpha} \rangle \neq 0$ (and so necessarily
$\langle e^{2i \theta} \rangle \neq 0$), to a paramagnetic state
with bond order which has $\langle \Phi_{\alpha} \rangle = 0$ but
still $\langle e^{2i \theta} \rangle \neq 0$. We can assume that
the background bond order is quenched, and hence need only
consider a theory of $n_{\alpha}$ fluctuations in this
environment.

For insulating systems, this theory will have essentially the same
structure as that in Section~\ref{sec:mottberry}, with the
difference that the coupling constants will now acquire a
modulation induced by the background bond order. The Berry phases
will again attempt to induce bond order in the regime with strong
fluctuations of $n_{\alpha}$, but this may be a redundant
effort--the bond order is already present in the underlying
Hamiltonian for the spins. As long as the underlying bond order
has an even number of spins per unit cell, the upshot is that we
can safely neglect the Berry phases, and the theory is simply that
of a $n_{\alpha}$ field undergoing a `quantum disordering'
transition in a partition function with positive weights and
short-range interactions: the critical theory for this is the O(3)
$\varphi^4$ field theory in 2+1 spacetime dimensions.

For systems with mobile charge carriers, we have to consider
possible long-range couplings induced by other gapless
excitations. These effects are quite strong and relevant in
metals, but are relatively unimportant in the superconducting
states considered in Section~\ref{sec:dopeberry}: superconductors
have low energy fermionic excitations at only select points in the
Brillouin zone, and these usually couple quite weakly to the
critical spin fluctuations \cite{vzs00}.

Having disposed of this simple possibility for the loss of
collinear magnetic order, let us now explore more fully the
possible phases that are allowed in a general interplay of
collinear magnetic and bond order parameters. The list of
possibilities is very large, but interplay with experiments should
help narrow the range of possibilities
\cite{zds02,zaanen02,zaanen02a}. As in Section~\ref{sec:eff} a
global view of the phase diagram is obtained most easily by
constructing an effective action for a $Z_2$ gauge theory. The
gauge theory must now be formulated for an order parameter on the
manifold $(S_2 \times S_1)/Z_2$, rather than $S_3/Z_2$, but the
strategy is quite similar to that of Section~\ref{sec:eff}. We
take the physical field $\Phi_{\alpha}$, and split it apart into
its $n_{\alpha}$ and $e^{i \theta}$ constituents; the overall sign
for the constituents involves a $Z_2$ gauge choice which can be
made differently at distinct points in spacetime. We can
compensate for this ambiguity by introducing a $Z_2$ gauge field,
$\sigma_{ij}$ (note that the physical interpretation of this
$\sigma_{ij}$ is entirely different from that in
Section~\ref{sec:eff}), and so obtain the effective partition
function \cite{zds02}:
\begin{eqnarray}
\mathcal{Z}_{\rm coll} &=& \int \prod_j d \theta_{j} d n_{j\alpha}
\delta \left( n_{j\alpha}^2 - 1 \right) \sum_{\{\sigma_{ij} = \pm
1\}} \exp
\left(-\mathcal{S}_{\rm coll} \right) \nonumber \\
 \mathcal{S}_{\rm coll} &=& -J_3
\sum_{\langle ij \rangle}   \sigma_{ij} n_{i \alpha} n_{j \alpha}
-J_4 \sum_{\langle ij \rangle}   \sigma_{ij} \cos(\theta_i -
\theta_j) - K \sum_{\Box} \prod_{\langle ij \rangle \in \Box}
\sigma_{ij}. \label{sz3}
\end{eqnarray}
For commensurate values of ${\bf K}$ there will be an additional
on-site anisotropy field arising from $V_{\rm comm}$ in
(\ref{comm}) $\sim - w \sum_j \cos (2p\theta_j)$.

\begin{figure}
\centerline{\includegraphics[width=3in]{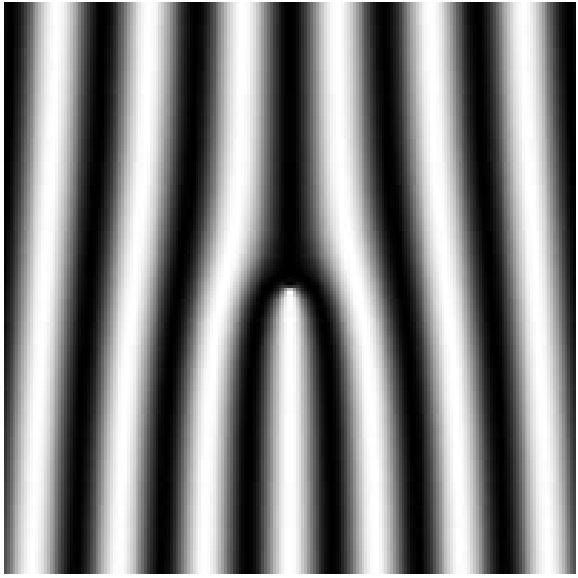}} \caption{A
stripe dislocation: gray scale plot of the bond order $Q_{\bf a}
({\bf r}) \sim \cos({\bf K}_x \cdot (2{\bf r}+{\bf a}) +
2\theta_x)$ (see (\protect\ref{theta}), (\ref{e1}) and
(\protect\ref{ab})) with $\theta_x = \pi/2+ (1/2)\tan^{-1} (y/x)$
containing a half vortex. The collective mode fractionalization
transition is between phases with and without suppression of these
defects. Both phases have neither magnetic or bond order.}
\label{fig5}
\end{figure}

What is the interpretation of a $Z_2$ gauge flux now ? It locates
the position of defects often called ``stripe dislocations'',
sketched in Fig~\ref{fig5}.
To see this, consider, as in Section~\ref{sec:eff}, the defects
associated with $\pi_1 ((S_2 \times S_1)/Z_2)$ \cite{zds02}. These
include ordinary vortices in the field $\theta$, under which
$\theta$ winds by an integer multiple of $2\pi$ upon encircling
the vortex. However, an additional class of {\em half-vortices}
are also allowed for which the $Z_2$ quotient plays a crucial
role: in these, $\theta$ winds by an half-odd-integer multiple of
$2 \pi$. At the same time, there is a corresponding trajectory for
$n_{\alpha}$ which connects polar opposite points on $S_2$; this
ensures that the physical field $\Phi_{\alpha}$ is single-valued
everywhere. In the context of (\ref{sz3}), we observe that, as in
Section~\ref{sec:eff} and Fig~\ref{fig1}, we will have
$\sigma_{ij}=-1$ only between the initial and final points of the
loop, to eliminate the branch cuts in $e^{i \theta}$ and
$n_{\alpha}$. Hence each such half vortex carries a $Z_2$ gauge
flux.

The phase diagram of $\mathcal{S}_{\rm coll}$ in the
three-dimensional $J_{3,4},K$ space is quite complex, but many of
its features can be understood by analyses of various
two-dimensional sections, most of which bear some formal
similarity to the model analyzed in Section~\ref{sec:eff}. We do
not want to run through the plethora of possibilities here, and
will merely focus on a particular phase transition
\cite{zaanen02,zds02} which has some intriguing experimental
implications.

Consider the region of small $J_{3,4}$ where $\langle
\Phi_{\alpha} \rangle = 0$ and $\langle e^{2i\theta} \rangle = 0$,
and so there is no magnetic or bond order. Let us first take a
small value of $K$---in the cuprates, we would imagine this is in
the overdoped region; then the field $\sigma_{ij}$ will fluctuate
strongly, and the stripe dislocations will proliferate. Such a
state has no vestiges of either magnetic or bond order, and is
therefore an ordinary superconductor or Fermi liquid. The strong
$\sigma_{ij}$ fluctuations also `bind' the $n_{\alpha}$ and $e^{i
\theta}$ fields to each other in the $\Phi_{\alpha}$ field, and
the latter constitutes a collective elementary excitation which
carries both spin and bond correlations; indeed, $\Phi_{\alpha}$
is simply an ordinary collective particle-hole excitation which
can be described by a time-dependent Hartree-Fock theory; in a
superconductor, this excitation could also avoid damping from the
particle-hole continuum, and so become a sharp quasiparticle. This
collective mode has 6 real components, corresponding to the 3
complex numbers required to define $\Phi_{\alpha}$ \cite{psvd01}.

Now increase the value of $K$; in the cuprates, this would
correspond to moving towards the underdoped region. This will
suppress the stripe dislocations until eventually there is a $Z_2$
gauge theory transition to a deconfined phase; this transition is
the analog of the transition across the line MC in Fig~\ref{fig2}.
The deconfined phase has dislocations suppressed, and unbound
quanta of $n_{\alpha}$ and $e^{i \theta}$. However it is still
true that $\langle \Phi_{\alpha} \rangle = 0$ and $\langle
e^{2i\theta} \rangle = 0$, and so there is no magnetic or bond
order. So this is a fractionalized phase in which quanta of
$\Phi_{\alpha}$ have separated into 3 real quanta of $n_{\alpha}$
and 2 real quanta of $e^{i \theta}$ {\em i.e.\/} the collective
mode has fractionalized. Note that unlike the fractionalized phase
of Section~\ref{sec:eff}, there are no neutral $S=1/2$
excitations, and the electron always retains its bare quantum
numbers.

Why is this $Z_2$ gauge fractionalization transition interesting ?
First, unlike that in Section~\ref{sec:nc}, it is based on
correlations (collinear magnetic and bond) that are actually
observed in the lightly doped cuprates. While long-range order
with $\langle \Phi_{\alpha} \rangle \neq 0$ and $ \langle e^{2 i
\theta} \rangle \neq 0$ is observed at low doping in some of the
cuprates, it is quite clear that all such conventional order has
essentially disappeared by the time we reach optimal doping.
However, remnants of such order may still be present in that the
stripe dislocations of Fig~\ref{fig5} have not yet proliferated.
This then opens up the possibility \cite{zds02,zaanen02} of a
quantum critical point near optimal doping
\cite{valla99,tallon01}, associated with the proliferation of
stripe dislocation with increasing doping, which is in the
universality class of the 2+1 dimensional $Z_2$ gauge theory. As
an added bonus, it has been recently argued \cite{sm02} that this
gauge theory remains strongly coupled even in the presence of
Fermi surface damping, and so is an attractive candidate for
anomalous behavior at higher temperatures.

\section{Conclusions}
\label{sec:conc}

This article has given a broad overview of phases of strongly
interacting electrons, going beyond those found in the
conventional Bloch-BCS theory of solid state physics. A plethora
of exotic and conventional states have been proposed in the
literature, and we attempted to bring some `order' into the
subject, by presenting a classification based on the physics of
Mott insulators, and by making connections between the
correlations in different states.

Some of the states considered here have conventional order
parameters, and could, in principle, also be obtained from a
weak-coupling instability of the Hartree-Fock/BCS theory of the
Fermi liquid or superconductor. These include the states with
collinear or non-collinear magnetic order, and with bond order. We
have chosen here, instead, to motivate them from the states of the
Mott insulator. Our approach properly accounts for the strongest
energy scales at the outset (the local repulsion between nearby
electrons), and so has a better handle on the energy scales of
various excitations. The Hartree-Fock/BCS theory could obtain
states with the same overall symmetry, but would give very
different estimates of the relative stability and energies of
various excitations. The selection of the wavevector, ${\bf K}$,
would also be based upon the Fermi surface geometry. In contrast,
the Mott-insulator-based approach used here, selects wavevectors
from very different criteria: {\em e.g.} in the bond ordered
states, it is the spin Berry phases and the resonance and
alignment of singlet valence bonds which selects the state, and
leads to predictions for the values of ${\bf K}$ \cite{sr91,vs99}
which have little to do with the Fermi surface geormetry. These
latter values appear to be consistent with observations
\cite{mexico}.

Another value in starting with the Mott insulator is that it also
allows a systematic classification of states with more exotic
order parameters, which cannot be expressed in terms of local
correlation functions. They appeared here as `quantum disordered'
states of different types of magnetic order; these `disordered'
states however possessed a topological order associated with the
suppression of certain defects which were classified by the global
topology of the magnetically ordered state. We also found a
description of the dynamics of this topological order in two
distinct $Z_2$ gauge theories, in systems with non-collinear and
collinear spin correlations respectively.

With some understanding of global features of the phase diagram at
hand, we can proceed with a study of possible quantum critical
points, and their influence on crossovers at finite temperatures
\cite{book}. For the cuprates, these issues have been reviewed in
Ref.~\cite{rmp}; the `stripe fractionalization' transition of
Section~\ref{sec:efrac} is only poorly understood, and much
additional work is required to understand its observable
consequences. The most intensive investigation of quantum
criticality has so far occurred in the heavy fermion compounds,
and most observations do not agree with the weak-coupling spin
density wave picture. Here, we think that the small Fermi surface
state of Section~\ref{sec:small} offers the prospect of leading to
many interesting related states, and of non-trivial quantum
critical points between them---this is an exciting avenue for
further research.

\subsection*{Acknowledgements}

I thank Eugene Demler, T. Senthil, and Matthias Vojta for recent
collaborations upon which this review is based. This research was
supported by US NSF Grant DMR 0098226.

\bibliographystyle{elsart-num}
\bibliography{mott}

\end{document}